\documentclass[twocolumn]{aa}

\def\kms{$\rm km\;s^{-1}$}
\def\kmsmpc{$\rm km\;s^{-1}\;Mpc^{-1}$}
\def\es{$\rm erg\;s^{-1}$}
\def\esc{$\rm erg\;s^{-1}\;cm^{-2}$}
\def\deg{$^\circ$}
\def\arcsec{$''$}
\def\arcmin{$'$}
\def\hour{$^{\rm h}$}
\def\min{$^{\rm m}$}

\def\ha{H$\alpha$}
\def\h2{H$_{2}$}
\def\hi{H~{\scriptsize I}}
\def\hii{H~{\scriptsize II}}
\def\nii{[N~{\scriptsize II}]}
\def\np{[N~{\scriptsize II}]$\,\lambda654.80$}
\def\ng{[N~{\scriptsize II}]$\,\lambda658.34$}
\def\sii{[S~{\scriptsize II}]}
\def\sp{[S~{\scriptsize II}]$\,\lambda671.65$}
\def\sg{[S~{\scriptsize II}]$\,\lambda673.08$}
\def\msun{M$_{\odot}$}
\def\lsun{$\rm L_{B_{\odot}}$}
\def\my{$\rm M_{\odot}\;yr^{-1}$}
\def\lha{$L_{\rm H\alpha}$}
\def\fha{$F_{\rm H\alpha}$}
\def\fhn{$F_{\rm H\alpha+[N~{\tiny II}]}$}

\begin{document}

\thesaurus{
	       03(11.09.1: NGC 3593; 
		  11.09.2;           
	          11.09.4;           
		  11.19.2;           
		  11.19.6)}          

\title{The Circumnuclear Ring of Ionized Gas in NGC~3593
\thanks{Based on observations carried out at ESO, La Silla (Chile) (ESO N. 52,
1-020) and on observations with the VATT: the Alice P. Lennon Telescope and the
Thomas J. Bannan Astrophysics Facility.}} 


\author{
	E.M.~Corsini        \inst{1}, 
	A.~Pizzella         \inst{2},
	J.G.~Funes,~S.J.    \inst{3},  
	J.C. Vega Beltr\'an \inst{4},
	and F.~Bertola      \inst{3}
}
	
\offprints{E.M. Corsini}
\mail{{\tt corsini@pd.astro.it}}

\institute{
Osservatorio Astrofisico di Asiago,
Dipartimento di Astronomia, Universit{\`a} di Padova, 
via dell'Osservatorio~8, I-36012 Asiago, Italy \and
European Southern Observatory, Alonso de Cordova 3107,   
Casilla 19001, Santiago 19, Chile \and
Dipartimento di Astronomia, Universit{\`a} di Padova,
vicolo dell'Osservatorio~5, I-35122 Padova, Italy \and
Osservatorio Astronomico di Padova, Telescopio Nazionale Galileo, 
vicolo dell'Osservatorio~5, I-35122 Padova, Italy}

\date{Received..................; accepted...................}

\authorrunning{Corsini et al.} 
\titlerunning{The Circumnuclear Ring of Ionized Gas in NGC~3593}

\maketitle

\begin{abstract}

We present the results of narrow-band \ha$+$\nii\ imaging of the
early-type spiral NGC~3593 in combination with a study of the flux
radial profiles of the \nii\ ($\lambda\lambda$ 654.80, 658.34 nm),
\ha, and \sii\ ($\lambda\lambda$ 671.65, 673.08 nm) emission lines
along its major axis.  The galaxy is known to contain two
counterrotating stellar discs of different size and luminosity. We
find that the \ha\ emission mainly derives from a small central region
of 57\arcsec$\times$25\arcsec . It consists of a filamentary pattern
with a central ring. This has a diameter of about 17\arcsec\
($\sim0.6\,h^{-1}$ kpc) and it contributes about half of the total
\ha\ flux. The ring is interpreted as the result of the interaction
between the acquired retrograde gas which later formed the smaller
counterrotating stellar disc and the pre-existing prograde gas of the
galaxy.


\end{abstract}

\section{Introduction}

By the mid-1980's the phenomenon of counterrotation had be\-en observed
in a significant number of elliptical and lenticular galaxies (see the
recent review by Bertola \& Corsini 1998). Only more recently the
presence of gaseous and/or stellar discs counterrotating with respect
to the main stellar body has been discovered in some early-type
spirals.  These galaxies are the Sab(s) \object{NGC~4826} (Braun et
al. 1992, 1994; Rubin 1994; Walterbos et al. 1994; Rix et al. 1995),
the Sb(r) \object{NGC~7217} (Merrifield \& Kuijken 1994), the Sa
\object{NGC~3626} (Ciri et al. 1995), the Sa pec \object{NGC~3593}
(Bertola et al. 1996), and the Sa(r) pec \object{NGC~4138} (Jore et
al. 1996).

This discovery poses new questions for the formation of disc galaxies.
It is almost impossible for an isolated galaxy to generate an internal
counterrotating component. In the general case, it is necessary to
assume that the galaxy has acquired material from outside. The infall
of external gas and the merging with gas-rich dwarf companions have
been recently investigated by means of numerical simulations (Thakar
\& Ryden 1996; Thakar et al. 1997) as possible mechanisms for
producing massive counterrotating gaseous discs in spirals.  The
presence of stellar counterrotation is usually interpreted as the end
result of an initial counterrotation of gas.  But gas acquisition in
spirals, more than in elliptical or in S0 galaxies, raises the problem
of the interaction between the accreted counterrotating and the
pre-existing corotating gas (e.g., Lovelace \& Chou 1996).
Recently some attempts have been made to explain special cases of
stellar counterrotation as due to a self induced phenomenon in
non-axisymmetric potentials (Evans \& Collett 1994; Athanassoula 1996;
Wozniak \& Pfenniger 1997).

NGC~3593 is an highly-inclined early-type spiral classified Sa pec by
Sandage \& Tammann (1981) and S0/a(s) uncertain by de Vaucouleurs et
al. (1991). Sandage \& Bedke (1994) in The Carnegie Atlas of Galaxies
(hereafter CAG), described NGC~3593 as characterized by a spiral dust
pattern made of patches rather than lanes. Such a dust pattern is
visible via its silhouette throughout the face of the disc (see panel
76 in CAG) obscuring the galaxy central regions to the north of the
major axis.  An overview of the optical properties of NGC~3593 is
given in Table~\ref{tab1}.

\begin{table}[ht]
\caption{Optical properties of NGC~3593}
\begin{flushleft}
\begin{tabular}{lc}
\hline
\noalign{\smallskip} 
parameter & value \\
\noalign{\smallskip} 
\hline
\noalign{\smallskip} 
Name                              &  NGC~3593 (UGC~6272)\\
Morphological type                &  Sa pec$^{\rm a}$; SAS0$\ast$$^{\rm b}$ \\ 
Position (equinox 2000.0)$^{\rm b}$  & \\
\hspace{.2truecm} right ascension $\alpha$ & 11\hour$\;$14\min$\;36\fs1$ \\
\hspace{.2truecm} declination     $\delta$ & $+12$\deg$\;$49\arcmin$\;07''$\\
Heliocentric systemic velocity $cz$$^{\rm c}$  & $612\pm10$ \kms \\
Position angle P.A.$^{\rm b}$                   &  92\deg \\
Isophotal diameters D$_{25}\times$d$_{25}$$^{\rm b}$ & 
$5\farcm25\times1\farcm95$ \\
Inclination $i^{\rm d}$                         &  67\deg  \\
Total corrected $B$ magnitude $B_T^0$$^{\rm b}$ &  $11.50$ mag \\
\noalign{\smallskip} 
\hline
\end{tabular}
\begin{list}{}{}
\item[$^{\rm a}$] from Sandage \& Tammann (1981)
\item[$^{\rm b}$] from de Vaucouleurs et al. (1991)
\item[$^{\rm c}$] from Bertola et al. (1996) 
\item[$^{\rm d}$] from Rubin et al. (1985)    
\end{list}
\end{flushleft}
\label{tab1}
\end{table}

In NGC~3593 Bertola et al. (1996) found two stellar exponential discs
of different scale lengths and surface brightnesses counterrotating
with respect to one other.  The smaller and less massive stellar disc
(disc 2) corotates with the \hi, the \h2\ and the ionized gas disc,
and counterrotates with respect to the larger and more massive stellar
disc (disc 1).  They also noticed that the radial intensity profile
for the gas emission lines has two maxima at about 9\arcsec\ to each
side of the centre.  The derived properties of the stellar, gaseous
and dust components of NGC~3593 are summarized in Table~\ref{tab2}.
They have been scaled to a distance of $7\;h^{-1}$ Mpc (Wiklind \&
Henkel 1992) with $H_{0} = 100\;h$ \kms\ Mpc$^{-1}$.

\begin{table}[ht]
\caption{Derived properties of NGC~3593}
\begin{flushleft}
\begin{tabular}{lc}
\hline
\noalign{\smallskip} 
parameter & value \\
\noalign{\smallskip} 
\hline
\noalign{\smallskip} 
Distance $d$$\,^{\rm a}$                  & 7 $h^{-1}$ Mpc \\
Total $B$ luminosity $L_B^0$              & $1.9\cdot10^{9}\;h^{-2}$ \lsun\\
Primary stellar disc (disc 1)$^{\rm b}$ & \\
\hspace{.2truecm} scale length $r_1$                     &  40\arcsec \\
\hspace{.2truecm} central surface brightness $\mu_{r,1}$ & 
19.9 mag arcsec$^{-2}$\\
\hspace{.2truecm} $B$ luminosity $L_{B_1}^0$ &$1.6\cdot10^{9}\;h^{-2}$ \lsun\\
\hspace{.2truecm} mass $M_1$               &$6.8\cdot 10^{9}\;h^{-1}$ \msun\\
Secondary stellar disc (disc 2)$^{\rm b}$ & \\
\hspace{.2truecm} scale length $r_2$                     &  10\arcsec \\
\hspace{.2truecm} central surface brightness $\mu_{r,2}$ & 
18.5 mag arcsec$^{-2}$\\
\hspace{.2truecm} $B$ luminosity $L_{B_2}^0$ &$3.5\cdot10^{8}\;h^{-2}$ \lsun\\
\hspace{.2truecm} mass $M_2$                & $1.5\cdot10^{9}\;h^{-1}$ \msun\\
Gas and dust & \\
\hspace{.2truecm} mass $M_{\rm HII}$$^{\rm c}$&$4\cdot10^{5}\;h^{-2}$ \msun\\
\hspace{.2truecm} mass $M_{\rm HI}$$^{\rm d}$ &$1.8\cdot10^{8}\;h^{-2}$ \msun\\
\hspace{.2truecm} mass $M_{\rm H_2}$$^{\rm a}$&$4.5\cdot10^{8}\;h^{-2}$ \msun\\
\hspace{.2truecm} mass $M_{\rm dust}$$^{\rm e}$&$5.1\cdot10^{5}\;h^{-2}$\msun\\
\noalign{\smallskip} 
\hline
\end{tabular}
\begin{list}{}{}
\item[$^{\rm a}$] from Wiklind \& Henkel (1992)
\item[$^{\rm b}$] from Bertola et al. (1996)
\item[$^{\rm c}$] from this paper
\item[$^{\rm d}$] from Krumm \& Salpeter (1979)  
\item[$^{\rm e}$] from Young et al. (1996)  
\end{list}
\end{flushleft}
\label{tab2}
\end{table}

The main purpose of this paper is to present a new narrow-band
\ha$+$\nii\ imaging of NGC~3593 and the flux radial profiles of the
\nii\ ($\lambda\lambda$ 654.80, 658.34 nm), \ha, and \sii\ 
($\lambda\lambda$ 671.65, 673.08 nm) emission lines measured along its
major axis. We use our photometric and spectroscopic data to confirm
the presence of a ring of ionized gas in the circumnuclear region of
NGC~3593.  It was previously suggested by Hunter et al. (1989) and
Pogge \& Eskridge (1993).

\section{Observations, data reduction and results}

\subsection{Narrow-band imaging}

The narrow-band \ha\ imaging of NGC~3593 was performed on March 8,
1997 at the 1.8~m Vatican Advanced Technology Telescope (VATT)
operated in the Mt. Graham International Observatory. A back
illuminated 2048$\times$2048 Loral CCD with 15 $\mu$m pixels was used
as detector at the aplanatic Gregorian focus, f/9. It yielded a field
of view of $6\farcm4\times6\farcm4$ with an image scale of $0\farcs4$
pixel$^{-1}$ after a $2\times2$ on-line pixel binning.  The gain and
the readout noise were 1.4$\rm \,e^{-}\,ADU^{-1}$ and 6.5$\rm \,e^-$
respectively.
 
We obtained $6\times10$ minutes emission-band images and $6\times10$
minutes continuum-band images using two narrow-band filters kindly
provided by R.C. Kennicutt.  The emis\-sion-band images were taken with
an interference filter ($\lambda_c = 658.0$ nm; $\Delta\lambda_{\rm
FWHM} = 7.0$ nm) isolating the spectral region characterized by the
redshifted \ha\ and \nii\ ($\lambda\lambda$654.80, 658.34 nm) emission 
lines. The continuum-ban\-d images were taken through an interference
filter ($\lambda_c = 645.0$ nm; $\Delta\lambda_{\rm FWHM} = 7.0$ nm),
which has been selected to observe an emission-free spectral region
sufficiently near to that of the emission-band filter in order to
subtract off the stellar continuum in the emission-band images. 
Different flat field exposures of the twilight sky were taken for each
of the two filters.

\begin{figure}
\vspace*{11cm}
\includegraphics{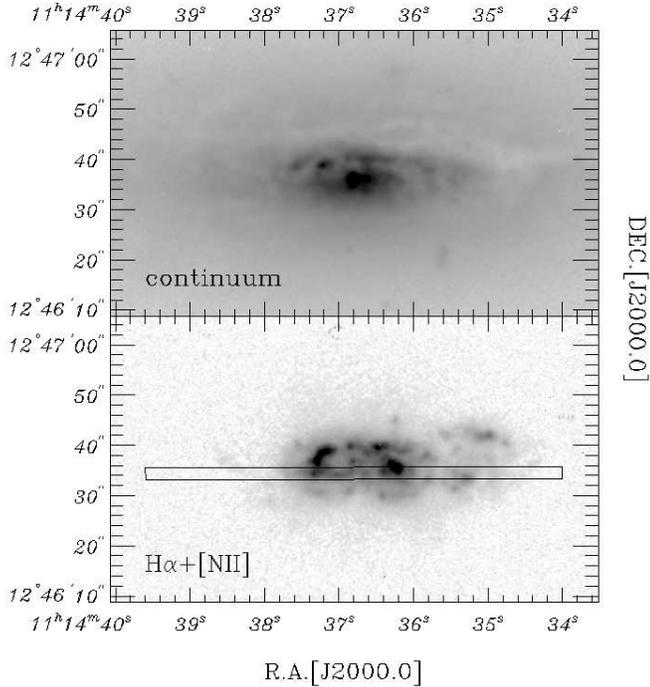}
\caption[]{The stellar continuum image (upper panel) and the 
continuum-subtacted \ha$+$\nii\ emission image (lower panel) of NGC~3593. 
The position of the slit of the ESO spectrum is shown in the lower panel.
The orientation of the images is north up and east to left. 
The astronomical coordinates were assigned to the images by measuring known 
positions of field stars on a Digitized Sky Survey image of NGC~3593}
\label{fig1}
\end{figure}

The data reduction was carried out using IRAF\footnote{IRAF is
distributed by the National Optical Astronomy Observatories which are
operated by the Association of Universities for Research in Astronomy
(AURA) under cooperative agreement with the National Science
Foundation}. The images were bias subtracted, flat field corrected
using the normalized flat field of the corresponding filter and then
sky subtracted. The sky level was determined by calculating the
average intensity in blank regions of the frames.  The images were
shifted and aligned to an accuracy of a few hundredths of a pixel
using the common field stars as a reference. Then they were averaged
(after checking that their PSF's were comparable) to obtain a single
continu\-um-band image (Fig.~\ref{fig1}, upper panel) and a single
emis\-sion-band image.  The cosmic rays were identified and removed
during the averaging routine. Gaussian fits to field stars in the two
final processed images yielded point spread function measurements of
$1\farcs1$ (FWHM). This corresponds to a physical resolution of 37
$h^{-1}$ pc at the assumed distance.  Finally the continuum-free image
of NGC~3593 (Fig.~\ref{fig1}, lower panel) showing the galaxy
\ha$+$\nii\ emission was obtained by subtracting the continuum-band
i\-mage, suitably sca\-led, from the emission-band image.  The mean
scale factor for the continuum image was estimated by comparing the
intensity of the field stars in the two bandpasses.

The flux calibration was performed using images of the optical
spectrophotometric standard star Feige 66 (Oke 1990) obtained during
the same night. The total \ha$+$\nii\ flux of NGC~3593 was measured by
summing the contribution of those pixels exceeding the $3\sigma$ sky
background noise level ($1\sigma=9.3\cdot10^{-17}$
\esc$\rm\;arcsec^{-2}$). The resulting value is \fhn\ $=
3.4\cdot10^{-12}$ \esc. The uncertainty on \fhn\ is estimated to be
$\sim20\%$. Our determination of the \ha$+$\nii\ flux is in agreement,
within the measurement errors, with that of Young et al. (1996).  We
subtract from this total emission flux the estimated contribution of
the two \nii\ lines to obtain the \ha\ emission flux for NGC~3593. At
the wavelengths of the redshifted \np , \ha ,and \ng\ lines the
transmission of the filter is $\sim66\%$, $\sim79\%$, and $\sim72\%$,
respectively. We know the \np/\ng\ ratio from the atomic transition
probabilities and we measured \ng/\ha$\sim0.46$ from our spectroscopical 
data (see \S~2.2). In fact the two \nii\ lines contribute $\sim18\%$ 
to the total emission in the conti\-nu\-um-free image. After this 
subtraction we find \fha\ $= 2.8\cdot10^{-12}$ \esc. Our \ha\ flux 
determination lies in between \fha\ $= 1.3\cdot10^{-12}$ \esc\ measured 
by Pogge \& Eskridge (1993) and \fha\ $= 8.6\cdot10^{-12}$ \esc\ by 
Hunter et al. (1989).  According to Burstein \& Heiles (1984) the
Galaxy contributes no appreciable foreground extinction for NGC~3593,
and noattempt was made to correct the computed flux for internal
extinction.  This flux corresponds to a total \ha\ luminosity of \lha\
$= 1.7\cdot10^{40}\;h^{-2}$ \es\ at the assumed distance.

\subsection{Long-slit spectroscopy}

For the study of radial profiles of the flux of the \nii , \ha\ and
\sii\ emission lines we adopted the same original spectral data of
NGC~3593 used by Bertola et al. (1996) to derive the kinematics of
stars and ionized gas.  They were obtained at the ESO 1.52 m
Spectroscopic Telescope in La Silla on February 18-19, 1994.  The
Boller \& Chivens Spectrograph was used in combination with the No.~26
1200 $\rm grooves\;mm^{-1}$ grating and a $2\farcs5\times4\farcm2$
slit.  Four major-axis (P.A. = 92\deg) spectra of 60 minutes were
taken with an instrumental resolution of $\sigma = 0.1$ nm (i.e. $\sim
48$ \kms\ at 620.0 nm). The spectral range was 520.0~--~719.0 nm. Each
NGC~3593 spectrum pixel corresponds to 0.097~nm$\times2\farcs43$ after
on-chip binning of 3 pixels along the spatial direction on the No.~24
FA2048L CCD. The details of the spectra reduction and calibration are
given in Bertola et al. (1996).

We measured the flux of the \nii\ lines ($\lambda\lambda$ 654.80,
658.34 nm), of the \ha\ line, and of the \sii\ lines ($\lambda\lambda$
671.65, 673.08 nm).  At each radius the line flux was obtained
separately for each emission line after fitting a polynomial to the
surrounding continuum, using the ESO-MIDAS\footnote{MIDAS is developed 
and maintained by the European Southern Observatory} package ALICE. At
some radii where the intensity of the emission lines was low, we
averaged adjacent spectral rows to improve their $S/N$ ratio.

\begin{figure}
\vspace*{10cm}
\includegraphics{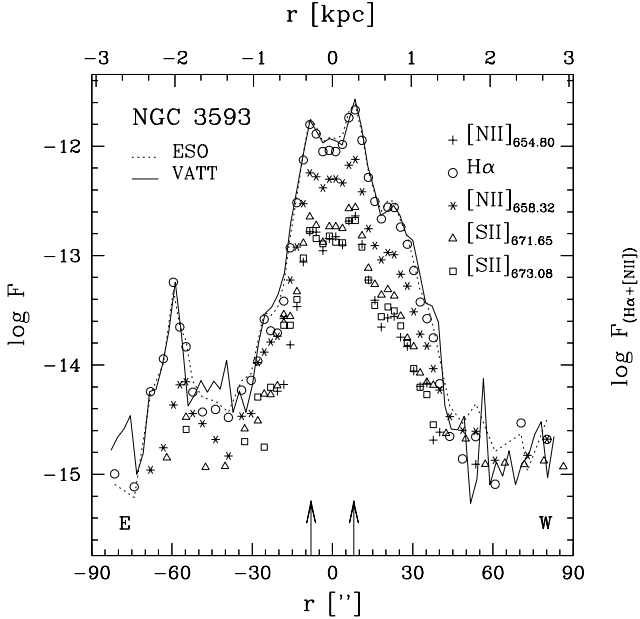}
\caption[]{The radial flux profile of the measured emission lines along
the major axis of NGC~3593. The {\it symbols\/} represent the values obtained 
for \ha\ ({\it circles\/}), the \nii\ lines ({\it crosses\/}
for $\lambda654.80$ nm and {\it asteriscs\/} for $\lambda658.34$ nm) and the
\sii\ lines ({\it triangles\/} for $\lambda671.65$ nm and {\it squares\/} for
$\lambda673.08$ nm) respectively. The radial flux profiles of the \ha$+$\nii\ 
emission from the ESO spectrum ({\it dotted line\/}) and from the VATT 
continuum-free image ({\it continuos line\/}) used in the flux calibration 
of the ESO spectrum are shown. In the VATT image several bad columns were 
present between $-70$\arcsec\ and $-50$\arcsec . The arrows indicate the 
position along the major axis of the molecular gas ring found by Wiklind 
\& Henkel (1992)}
\label{fig2}
\end{figure}

The flux calibration was performed by comparing the major axis
\ha$+$\nii\ emission radial profiles of the ESO spectrum with that of
VATT continuum-free image (see Fig.~\ref{fig2}).  The uncalibrated
profile of the ESO spectrum was obtained by summing, at each radius,
the contribution of the \ha\ and of the two \nii\ lines scaled for the
transmission of the VATT emission-band filter at their relevant
wavelengths. The VATT calibrated profile was obtained from the central
row of the continuum-free image. The image has been previously rotated
counterclockwise by 2\deg\ and rebinned in $6\times8$ pixels to
reproduce the scale, the slit width and the seeing of the ESO
spectrum.  Because of the differences in instrument, weather
conditions and set-up of the two observing runs this should be
considered only a rough but useful calibration.

The flux radial profiles of all the measured emission lines show the
same trend (Fig.~\ref{fig2}).  They are characterized by a central
relative minimum and two maxima at $r = \pm 8\farcs5$ . A secondary
peak in the fluxes of the emission lines has been detected at
$r\simeq+56$\arcsec .
For $-30$\arcsec\ $\leq r \leq +20$\arcsec\ the flux ratio between the
\ng\ and the \ha\ lines ranges between 0.35 and 0.56, the flux ratio
between the \sp\ and the \ha\ ranges between 0.10 and 0.16, and the flux
ratio between the \sg\ and the \sp\ lines is virtually constant at 0.76.

\section{Discussion and conclusions}

Our continuum-free image reveals that the \ha$+$\nii\ flux of NGC~3593
mainly derives from a small central region of about 
57\arcsec$\times$25\arcsec\ ($\sim2\times0.8\,h^{-2}$ kpc$^{-2}$) as
projected on the sky. It constits of a filamentary pattern dominated
by a ring and two bright \hii\ structures. The centre of the ring
coincides with the position of the maximum intensity in the
continuum. The two \hii\ structures are located to the west of the
centre at a distance of about 24\arcsec\ and about 30\arcsec\ at
position angles of 109\deg\ and 79\deg\ respectively. A faint
\hii\ complex is detected to the east at about 59\arcsec\ from the 
centre at a position angle of 94\deg . The area of the ring is about 
0.17 $h^{-2}\;{\rm kpc}^{-2}$ ($\sim11\%$ of the total line-emitting
region). It contributes about half of the total line flux. Fitting
this maximum-emission region, we derived for the ring an apparent
major-axis diameter $D = 17\farcs4\pm0\farcs8$ ($\sim0.6\,h^{-1}$
kpc), a position angle $\theta = 90$\deg$\pm5$\deg, and an ellipticity
$\epsilon = 0.62 \pm 0.08$ corresponding to an inclination $i =
68$\deg$\pm5$\deg . The derived value of the ring inclination is in
agreement within the errors with the galaxy inclination of 67\deg\
derived by Rubin et al. (1985) from the disc parameters.  The position
of the two maxima in the radial flux profiles of the all the measured
emission lines corresponds to the radii at which the ring intersects
the line of the nodes.
 
We have obtained the photometric and spectroscopic evidence for the
existence in NGC~3593 of a circumnuclear ring of ionized gas. It is
associated with molecular gas, which is also distributed in a nuclear
ring-like configuration as discovered by Wiklind \& Henkel (1992). The
centre of the CO ring coincides with the galaxy kinematical center and
its radius ($R_{\rm CO} \sim 7\farcs9-8\farcs8$) is located at the
turning point of the ionized gas rotation curve.

In the past years, Hunter et al. (1989) were not able to distinguish
between a small gaseous disc and a possibly ring-like structure due to
the lower resolution of their \ha\ image.  Pogge \& Eskridge (1993)
derived the same overall morphology and suggested the presence of the
ring.  The extent of the \ha$+$\nii\ emitting region is about
$44''\times18''$, as deduced from their published plate. In this image
the diffuse emission surrounding the structure of the ring is not
visible. They measured for the \ha\ ring the same major-axis position
angle ($\theta = 90$\deg$\pm2$\deg) and a smaller diameter ($D =
14\farcs4\pm1\farcs4$). This difference is probably due to the clumpy
morphology of the ionized gas emission. No estimate of the ring
ellipticity (or inclination) is given.

As it appears from the kinematics along the major axis of NGC~3593
(Bertola et al. 1996) the ionized gas ring corotates with the gaseous
and the secondary stellar discs and counterrotates with respect to the
primary stellar disc.  From Wiklind \& Henkel (1992) we derive that
the ionized and the molecular gas components corotate.
 
The \ha\ luminosity can be used to estimate the current star formation
rate (SFR) of NGC~3593. Following the SFR relation by 
Kennicutt (1983) we find for NGC~3593 a SFR$_{\rm H\alpha}$ (total)$ =
0.15\;h^{-2}$ \my. This value is about one tenth of the SFR derived
from the far infrared (FIR) luminosity by Wiklind \& Henkel
(1992). Indeed it is SFR$_{\rm FIR}$ (total)$ = 1.6\;h^{-2}$ \my.  The
observed \ha\ flux is about $9\%$ of the value needed to make the SFR
derived from the \ha\ luminosity comparable to that from the FIR
luminosity.  It means that NGC~3593 is not a transparent system. In
fact NGC~3593 is a dust-rich system as one can also infer by a
comparison of its dust-to-\hi\ gas mass ratio with the mean values
found by Bregman et al.  (1992) for the early-type disc galaxies
investigated by Roberts et al.  (1991).

We can estimate the mean past SFR from the mass and the age of the stellar 
disc. Following Kennicutt et al. (1994), but using our determination of 
the mass of stellar disc, we find $<{\rm SFR}>$ (past) $ = 1.4\;h^{-1}$ \my.
For NGC~3593 the birthrate parameter $b$ calculated as the ratio of current 
SFR (from FIR luminosity) to average past SFR is $b = 1.1\;h^{-1}$. 
For comparison, Kennicutt et al. (1994) found that this value for a sample of 
18 non-interacting and/or peculiar Sa's ranges from 0.01 to 0.1 
(with $h = 0.75$ \kmsmpc ). So the central part of NGC~3593 is characterized
by a value of $b$ which is high compared to the majority of early-type spirals.
The current SFR is slightly higher than the SFR in the recent past.
Given the uncertainty in estimation of the SFR the only tenable
conclusion is that it has been approximately constant.

The total ionized gas mass can be derived from the \ha\ luminosity
under the assumption of case B recombination (see Osterbrock 1989).
For a given electron temperature and density, the \hii\ mass can
be written:
\begin{equation}
M_{\rm HII} = \left( L_{\rm H\alpha} m_{\rm H} /N_{e} \right) /
\left( 4 \pi j_{\rm H\alpha} / N_{\rm e} N_{\rm p} \right)
\end{equation}
where \lha\ is the \ha\ luminosity, $m_{\rm H}$ is the mass of the
hydrogen atom, $j_{\rm H\alpha}$ is the \ha\ emissivity, and 
$N_{\rm e}$ and $N_{\rm p}$ are the electron and the proton densities.
The term $4 \pi j_{\rm H\alpha} / N_{\rm e} N_{\rm p}$ is insensitive
to changes of $N_{\rm e}$ over the range $10^2$--$10^6$ cm$^{-3}$.
It decreases by a factor 3 for changes of $T_{\rm e}$ over the range
$5\cdot10^3\,$\deg K--$2\cdot10^4\,$\deg K (Osterbrock 1989).
For an assumed temperature $T_{\rm e}=10^4\,$\deg K, the electron density
can be estimated to be $N_{\rm e}=10^3$ cm$^{-3}$ from the measured
\sii\ lines ratio. It results $M_{\rm HII} = 4\cdot10^4\;h^{-2}$
\msun\ for the observed \lha\ and a value ten times larger for the
\lha\ corrected for extinction. 

The presence of a counterrotating star-forming \ha\ ring in the
nuclear regions of a galaxy was found for the first time in NGC~4138
(Jore et al. 1996). The discovery of the same feature in NGC~3593 is
notable, as the two galaxies are very similar.  They are both
early-type spirals with a small bulge and an apparently undisturbed
morphology apart from the dust lanes crossing their stellar
components. But they each have two counterrotating stellar discs with
the more massive one containing about the $80\%$ of the stars. In both
galaxies a gaseous disc is present and it rotates in the same
direction as the less massive stellar disc. In the two objects the
total counterrotating mass amounts to about the $25\%$ of their
luminous mass.

The ring in NGC~3593 gives further significance to the numerical
results of Thakar et al. (1997) for NGC~4138. They adopted a
two-component galaxy model with a stellar gas-free exponential disc
and a truncated spherical isothermal halo to investigate the origin of
the massive counterrotating component.  In one of their simulations,
namely I6, they placed in the stellar disc a prograde ring of gas. In
this case the observed counterrotating \ha\ ring is the relic of the
collisions between the infalling retrograde gas clouds and the
pre-existing but less massive prograde gas. Such collisions could have
triggered the local enhancement of the stellar formation rate in the
region of the ring.  The same scenario can be applied to the case of
NGC~3593. The formation of one or more gaseous rings in
counterrotating systems as due to the two-stream instability has been
discussed by Lovelace et al. (1997).

As to the remaining spirals hosting counterrotation listed in \S~1,
in NGC~4826 the interaction between the newly supplied counterrotating
gas and the old corotating one is actually observed in the so-called
`transition region' (Rubin 1994). In NGC~3626 no gaseous ring has been
detected (Garc\'\i a-Burillo et al. 1998).  In NGC~7217 the
ionized gas rotates in the same direction as the primary stellar disc
(Merrifield \& Kuijken 1994). The presence of three gaseous rings has
suggested to Athanassoula (1996) an internal origin for its stellar
counterrotation as due to the decay of a bar.  In this sense gaseous
rings in this kind of objects could be useful to understand if the
origin of counterrotation is external or internal.  In the case of an
external origin a gaseous ring could be considered as the signature of
the presence of an original amount corotating gas.

\acknowledgements
We wish to thank G. Galletta for useful discussions and J.E. Beckman
for his valuable comments on the manuscript. We are most grateful to
R.C. Kennicutt for the use of its interference filters set at the
VATT, and to the Vatican Observatory Research Group for the allocation
of time for our observations.  In particular we thank R. Boyle, S.J
for his help during the observing run at the VATT. The research of AP
was partially supported by an {\em Acciaierie Beltrame\/} grant. JCVB
acknowledges the support by a grant of the Telescopio Nazionale
Galileo and the Osservatorio Astronomico di Padova. EMC thanks the
Instituto de Astrof\'{\i}sica de Canarias for hospitality while this
paper was in progress. This research has made use of The Digitized Sky
Survey, which was produced at the Space Telescope Science Institute
under U.S. Government grant NAG~W-2166.


\begin{thebibliography}{}
\bibitem[1996]{ath}  Athanassoula, E. 1996, In: `Barred Galaxies', IAU Coll. 
  117, Buta R., Crocker D.A., B. G. Elmegreen (eds.), ASP Conf. Ser. Vol. 91, 
  ASP, San Francisco, p. 309  
\bibitem[1998]{ber1}  Bertola, F., Corsini, E.M. 1998, In: `Galaxy 
  Interactions at Low and High Redshift', IAU Symp. 186, Sanders D.B. (ed.),
  ASP Conf. Ser., ASP, San Francisco, in press
\bibitem[1996]{ber2}  Bertola, F., Cinzano, P., Corsini, E.M., Pizzella, A., 
  Persic, M., Salucci, P. 1996, ApJ, 458, L67
\bibitem[1992]{bra1}  Braun, R., Walterbos, R.A.M., Kennicutt Jr., R.C. 
  1992, Nature, 360, 442
\bibitem[1994]{bra2}  Braun, R., Walterbos, R.A.M., Kennicutt Jr., R.C., 
  Tacconi, L.J. 1994, ApJ, 420, 558 
\bibitem[1992]{bre}  Bregman, J.N, Hogg, D.E., Roberts, M.S. 1992, ApJ, 387,
  484
\bibitem[1984]{bur}  Burstein, D., Heiles, C. 1984, ApJS, 54, 33
\bibitem[1995]{cir}  Ciri, R., Galletta, G., Bettoni, D. 1995, Nature, 375, 
  661  
\bibitem[1991]{dev}  de Vaucouleurs, G., de Vaucouleurs, A., Corwin Jr.,
  H.G., Buta, R.J., Paturel, G., Fouqu\`e, P. 1991, Third Reference Catalogue
  of Bright Galaxies, Springer-Verlag, New York  
\bibitem[1994]{eva}  Evans, N.W., Collett, J.L. 1994, ApJ, 420, L67
\bibitem[1998]{gar}  Garc\'\i a-Burillo, S., Sempere, M.J., Bettoni, D. 1998,
  ApJ, in press   
\bibitem[1989]{hun}  Hunter, D.A., Thronson Jr., H.A., Casey, S., Harper, D.A.
  1989, ApJ, 341, 697
\bibitem[1983]{ken1}  Kennicutt Jr., R.C. 1983, ApJ, 272, 54 
\bibitem[1994]{ken2}  Kennicutt Jr., R.C., Tamblyn, P., Congdon, C.W. 1994, 
  ApJ, 435, 22
\bibitem[1979]{kru} Krumm N., Salpeter, E.E., 1979, ApJ, 228, 64
\bibitem[1996]{jor}  Jore, K.P., Broeils, A.H., Haynes, M.P. 1996, AJ, 112, 
  438 
\bibitem[1996]{lov1}  Lovelace, R.V.E., Chou, T. 1996, ApJ, 468, L25
\bibitem[1997]{lov2}  Lovelace, R.V.E., Jore, K.P., Haynes, M.P. 1997, ApJ, 
  475, 83      
\bibitem[1994]{mer}  Merrifield, M.R., Kuijken, K. 1994, ApJ, 432, 575
\bibitem[1990]{oke}  Oke, J.B. 1990, AJ, 99, 1621
\bibitem[1989]{ost}  Osterbrock, D.E. 1989, Astrophysics of Gaseous Nebulae 
  and Active Galactic Nuclei, University Science Book, Mill Valley
\bibitem[1993]{pog}  Pogge, R.W., Eskridge, P.B. 1993, AJ, 106, 1405 
\bibitem[1995]{rix}  Rix, H.-W., Kennicutt Jr., R.C., Braun, R., Walterbos,
  R.A.M. 1995, ApJ, 438, 155
\bibitem[1991]{rob}  Roberts, M.S., Hogg, D.E., Bregman, J.N., Forman, W.R.,
  Jones, C. 1991, ApJS, 75, 751
\bibitem[1994]{rub1}  Rubin, V.C. 1994, AJ, 107, 173
\bibitem[1985]{rub2}  Rubin, V.C., Burstein, D., Ford Jr., W.K., Thonnard, N.
  1985, ApJ, 289, 81
\bibitem[1993]{sag}  Sage, L.J. 1993, A\&A, 272, 123
\bibitem[1994]{san1}  Sandage, A., Bedke, J. 1994, The Carnagie Atlas of 
  Galaxies, Carnagie Institution and Flintridge Foundation, Washington (CAG)
\bibitem[1987]{san2}  Sandage, A., Tammann, G.A. 1981, A Revised Shapley-Ames 
  Catalog of Bright Galaxies, Carnagie Institution, Washington 
\bibitem[1996]{tha1}  Thakar, A.R., Ryden, B.S. 1996, ApJ, 461, 55
\bibitem[1997]{tha2}  Thakar, A.R., Ryden, B.S., Jore, K.P., Broeils, A.H. 
  1997, ApJ, 479, 702
\bibitem[1994]{wal}  Walterbos, R.A.M., Braun, R., Kennicut Jr., R.C. 1994,
  AJ, 107, 184
\bibitem[1992]{wik}  Wiklind, T., Henkel, C. 1992, A\&A, 257, 437
\bibitem[1997]{woz}  Wozniak, H., Pfenniger, D. 1997, A\&A, 317, 14
\bibitem[1996]{you}  Young, J.S., Allen, L., Kenney, J.D.P., Lesser, A., 
   Rownd, B. 1996, AJ, 112, 1903 

\end{thebibliography}
\end{document}